\documentstyle[aps,epsf]{revtex}  

\begin{document}        

\baselineskip 14pt
\title{The ATLAS Pixel Project}
\author{John Richardson}
\address{Lawrence Berkeley Laboratory}
\author{representing the ATLAS Pixel Collaboration}  

\maketitle          

\begin{abstract}    
The ATLAS experiment at the Large Hadron Collider will incorporate
discrete, high-resolution tracking 
sub-systems in the form of segmented silicon detectors with 40MHz
radiation-hard readout electronics. 
In the region closest to the $pp$ interaction point,
the thin silicon tiles will be segmented into a pixel geometry providing
two-dimensional space-point information.
The current status of the ATLAS pixel project will be presented with an
emphasis on the performance of the front-end electronics and
prototype sensors.
\end{abstract}   	

\section{Introduction}             
The ATLAS experiment \cite{TP} at the CERN Large Hadron Collider
will function as an exploratory tool
for the Higgs boson along with possible new physics phenomena beyond the scope of the Standard Model such as supersymmetry.
An average of 23 minimum-bias events will be produced every 25ns at the
design luminosity ($\cal{L}$ = 10$^{34}$cm$^{-2}$s$^{-1}$). 
Hence interesting signatures will be buried amongst
extraordinary degrees of background.
In order to realise fully the physics potential of 14TeV $pp$ collisions at
this luminosity, ATLAS must be capable of high-efficiency track reconstruction
with excellent impact-parameter resolution for 3D-vertexing and b-tagging. 
Even at lower luminosities of 10$^{33}$cm$^{-2}$s$^{-1}$, ATLAS will encounter
millions of $t\bar{t}$ pairs per year for top-quark analyses along 
with a plethora of $b$-quarks
enabling the possibility of CP-violation studies in the $b$-sector.
For these areas of study, the tracking system must again 
deliver high spatial precision and reconstruction efficiency along with 
excellent momentum resolution. 

In addition to the performance requirements, 
the extremely harsh radiation environment defines perhaps the greatest
challenge of all to the detector elements in ATLAS. At a radius of
30cm in the barrel region we expect to encounter a hadronic 
fluence\footnote{expressed in terms of the 1MeV-neutron 
NIEL equivalent damage} of $\approx$ 
1.3$\times$10$^{14}n_{\mbox{eq}}$cm$^{-2}$ after 10 years of operation.
At 10cm the figure is closer to 1.0$\times$10$^{15}n_{\mbox{eq}}$cm$^{-2}$.

The Inner Detector of ATLAS \cite{TDR1} will incorporate high-resolution, low mass
discrete tracking components in the form of
silicon microstrip and silicon pixel \cite{TDR2} detectors. The pixel elements
will form the innermost layers by virtue of their higher granularity 
and because
they are more suited to providing pattern recognition capability at the
greatest track densities. Also, having a greater degree of segmentation
than their strip-geometry counterparts naturally endows them with lower 
noise-per-channel and greater immunity to radiation damage.

\section{Layout of the Pixel System}
\begin{figure}[tbp]	
\centerline{\epsfxsize 13 cm \epsfbox{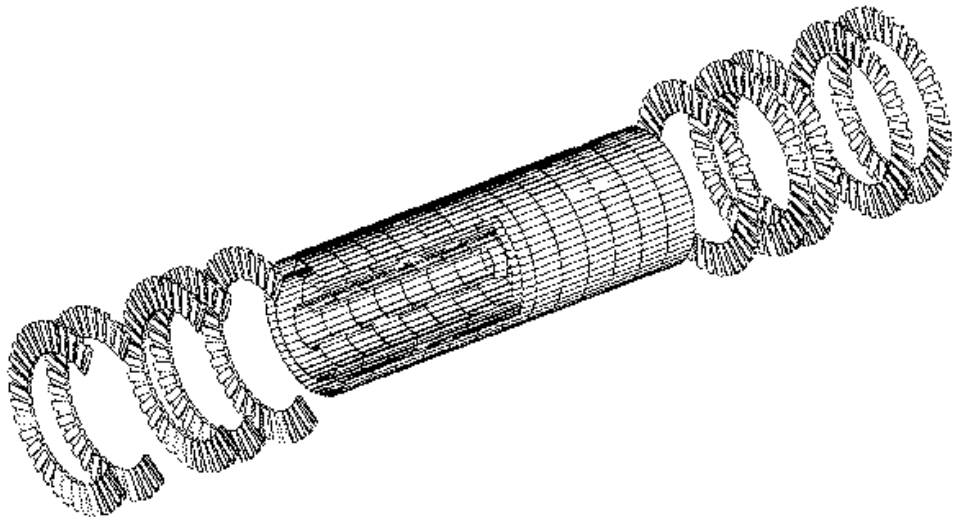}}   
\vskip -.2 cm
\caption{\it Positioning of the modules in the pixel system.}
\label{fi:tiles}
\end{figure}

Figure \ref{fi:tiles} shows the relative positioning of the individual 
modules within the pixel sub-detector.
There are 3 barrel layers at radii of 4.3, 10.1 and 13.2cm along with 
5 disc structures which extend to $\pm$78cm in the forward directions;
providing pseudorapidity coverage to 2.5.
The innermost barrel layer, (known as the $B$-layer), is not expected to 
survive the 10-year duration of the experiment and will therefore be replaced
periodically.
A single module comprises a silicon sensor {\it tile} with 16 front-end readout
chips bump-bonded to it along with one {\it Module Controller Chip} (MCC). 
The pixel detector will incorporate 2228 such modules with a 
total active area of 2.3m$^2$.

\section{Module structure and hybridization}
There are currently two technology options being pursued for 
module hybridization known as the {\it flex-hybrid} approach
and the {\it MCM-D} technique\footnote{{\bf M}ulti 
{\bf C}hip {\bf M}odule - {\bf D}irect.}.
The former incorporates a thin kapton hybrid which is glued to the sensor 
backplane to provide the necessary bussing between the FE-chips, MCC and
optical transmission components.
The front-end chips, which are bump-bonded to the opposite face of the
sensor, protrude along the long-edges of the module in order
that their 48 power,data and control bond-pads may be wire-bonded 
up to the hybrid.
The MCC sits on top of the hybrid along with the other
electrical components such as decoupling capacitors,
optical-link termination devices and temperature sensor.

The MCM-D method eliminates the need for wire bonds as the bussing is 
processed on to the sensor surface (active-side) in four
layers of copper metallisation. The long sensor dimension is extended 
to provide a region of `dead' silicon upon which the MCC and peripheral
components are mounted. The sensor is also wider in order that the
connections from the front-end chips to the power and control bus' may be formed using bump-bonds in the same
way that the individual pixel implants are connected to the preamp inputs.

Bump-bonding at 50$\mu$m pitch was until recent years considered
to be a serious issue.
However, there are several vendors with whom very promising 
experience has been gained recently in both the Indium and $PbSn$-solder
technologies with pre-flip-chip yields of order 10$^{-5}$.
For the Indium process, the flip-chip stage involves pressing together
the electronics chip and sensor, (both of which have the bumps grown on them), 
to form a cold weld. Solder bumps are grown only on the electronics surface
using a plating and re-flow technique. The recipient sensor surface
is plated with a wettable metallisation.
Flip-chipping in this case involves bringing the two surfaces into
accurately aligned contact and re-flowing the solder at 250$^{\mbox{o}}$C.

\section{Prototype Sensors}
The pixel sensors will be formed from
250$\mu$m-thick high-$\rho$ Silicon except within the $B$-layer
where it is hoped to utilise material as thin as 200$\mu$m.
High-$\rho$ silicon which begins as $n^-$-type inverts to $p$-type
after a hadronic dose of 
$\approx$ 2.5$\times$10$^{12}n_{\mbox{eq}}$cm$^{-2}$.
Beyond this dose the depletion voltage rises as the effective 
carrier concentration continues to increase.
After 10-years of radiation exposure, the required reverse-bias voltage
for full depletion of the pixel sensors will be in excess
of 1000V. 
Since the maximum operating voltage is specified to be 600V,
we will be forced to operate them in a partially-depleted
regime.
For this reason all of the prototype designs are based on
$n^+$-type pixel implants so that post-inversion, the junction
resides on the active side.

\begin{figure}[tbp]	
\centerline{
\epsfxsize 6.2cm 
\epsfysize 3.6cm \epsfbox{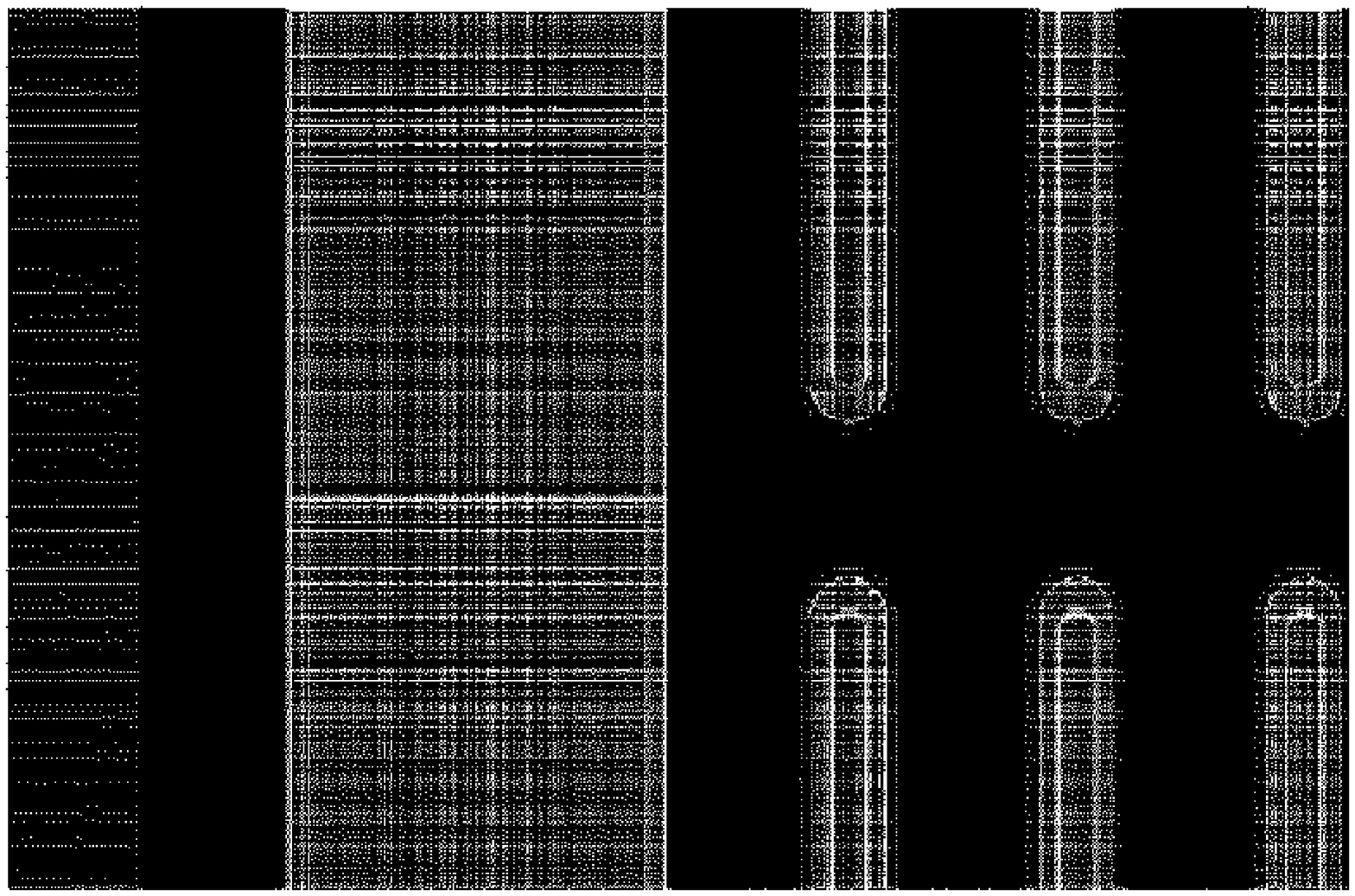}
\hspace{2mm}
\epsfxsize 6.2cm 
\epsfysize 3.6cm \epsfbox{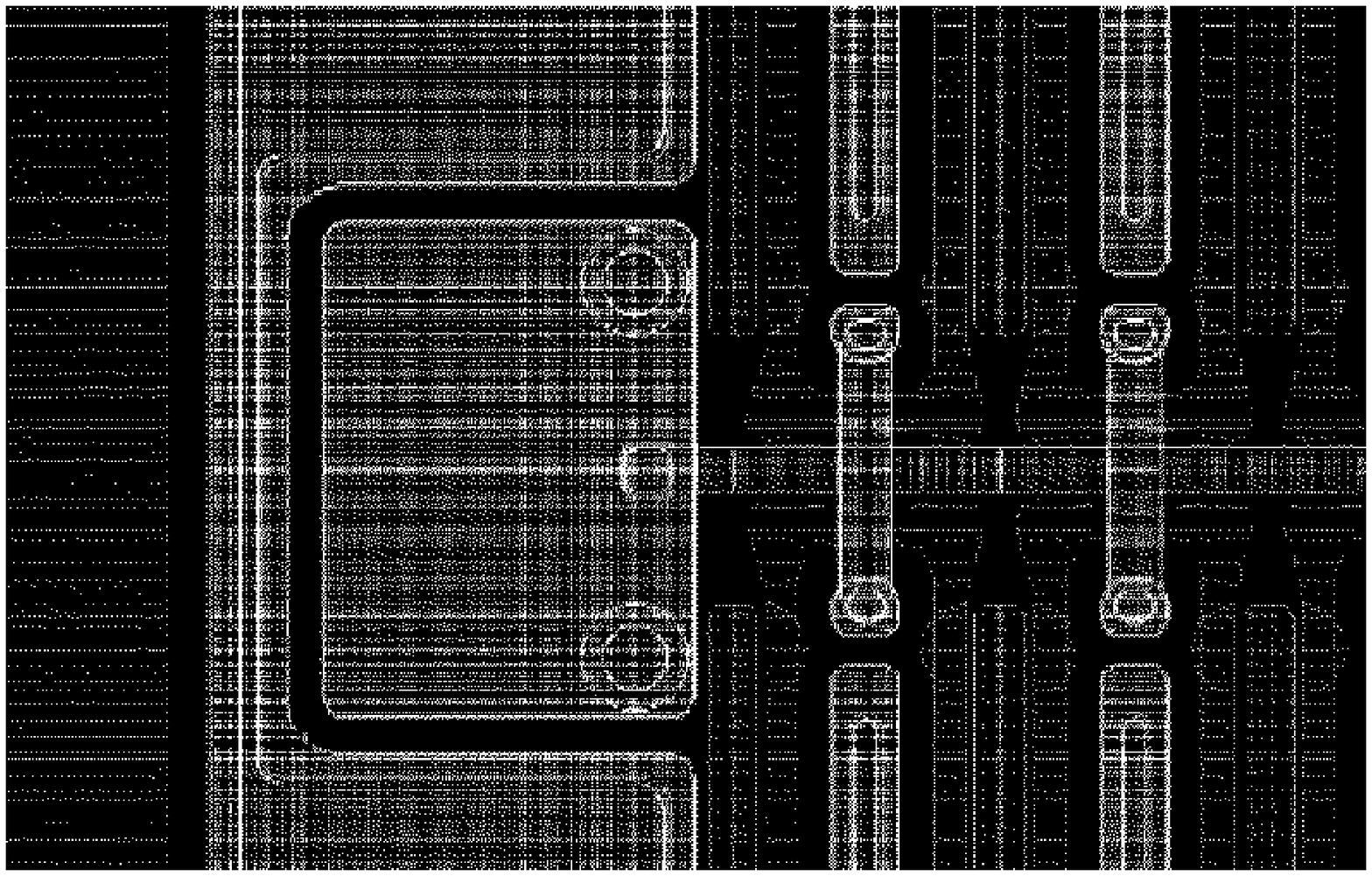}
\hspace{2mm}
\epsfxsize 6.2cm 
\epsfysize 3.6cm \epsfbox{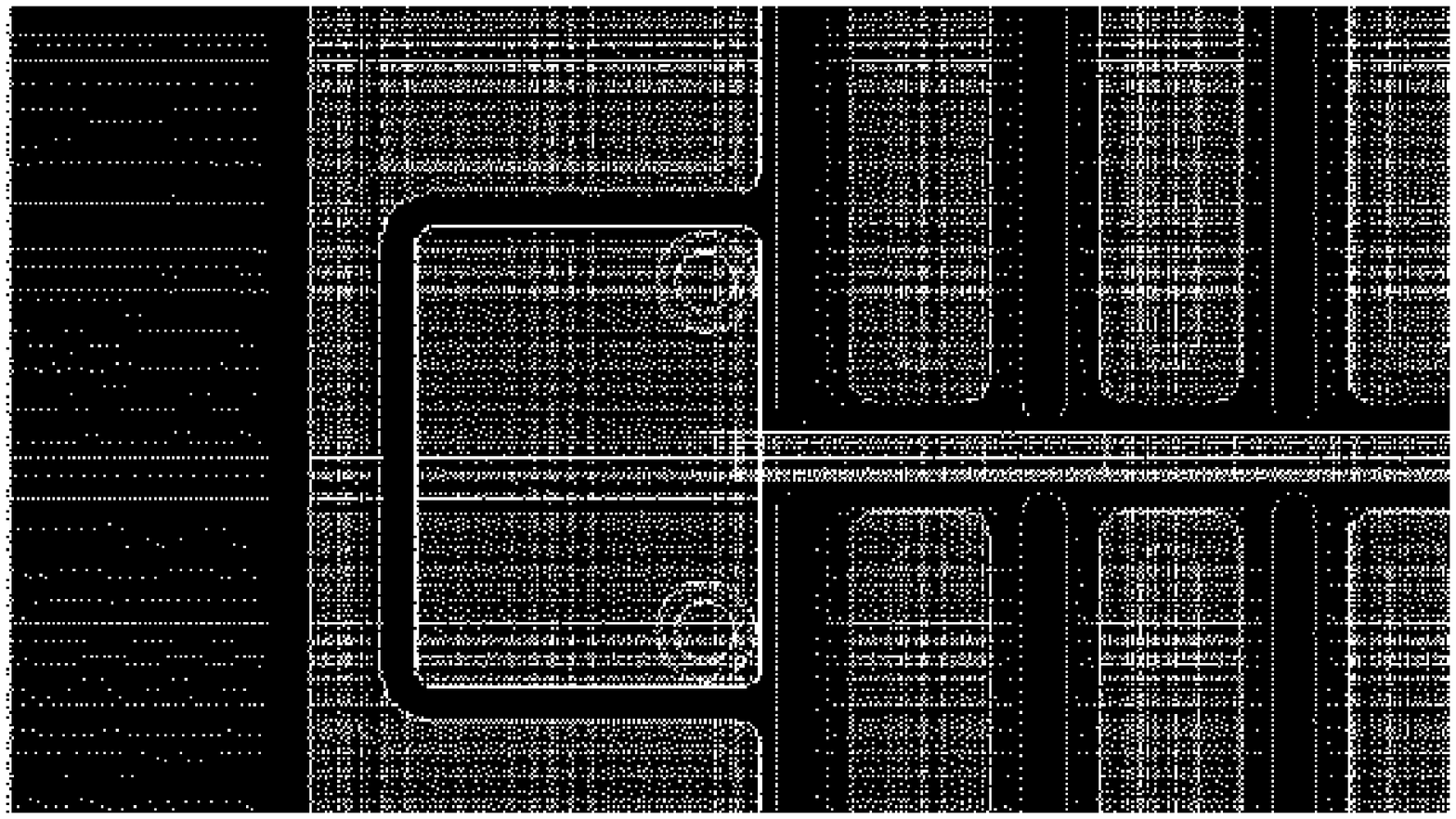}}
\vskip 0.5cm
\caption{\it The Tile-1 (left), Tile-2 (centre) and SSG sensor designs.}
\label{fi:st1st2}
\end{figure}

The first-phase of sensor prototyping was conducted with
two vendors; Seiko in Japan and C.I.S. in Germany.
Identical wafer designs were submitted to both and
featured two full-size tile designs.
One of these had {\it p-stop} isolation structures, 
(known as `Tile-1'), whilst
the other, (Tile-2), used the {\it p-spray} \cite{Tilman} technique
to over-compensate the electron
accumulation-layer which otherwise has the effect
of shorting $n^+$-implants together.
Along with these devices there were a variety of smaller
detector designs with a pixel-array corresponding to that of a single
readout chip.
Some of these matched the Tile-1 and Tile-2 designs (known as
ST1 and ST2 respectively).
Figure \ref{fi:st1st2} shows the layout of these two designs in detail.
The ST1 design is on the left and the $p^+$-type isolation rings
may be seen surrounding the $n^+$-pixel implants.
In the ST2 design there are also ring-like structures around each pixel
but these are now $n^+$-type and have the purpose of reducing the inter-pixel
capacitance.
The ST2 design also incorporates reach-through biasing which is not
possible to implement with p-stop isolation.
The third design shown in figure \ref{fi:st1st2} is known 
as SSG\footnote{{\bf S}ingle {\bf S}mall {\bf G}ap}.
This also utilises p-spray isolation but does not include 
any intermediate $n^+$ structures. Neighbouring pixels rather have just 
small gaps between them.

\section{Front-End Readout Electronics}
There are many demands on the front-end readout electronics.
Insensitivity to the changes in 
transistor thresholds and transconductance brought on
by a total ionising dose of 30MRad, (in layer-1 
after 10-years of operation), must be built into the design.
In addition, the front-ends must be able to deliver high-efficiency
performance when the operation of the sensors has been 
compromised by hadronic damage. 
This will involve leakage-current
tolerence to the level of 100nA/channel whilst maintaining
sensitivity to charge spectra much reduced by partial-depletion
operation and charge-trapping effects.
The power consumption must not exceed 40$\mu$W per channel,
the timewalk must be contained to the 25ns bunch-crossing period
of the LHC and the cross-coupling between neighbouring pixels
should account for a charge loss of no more than 5\%.

Over the last year and a half, the {\it demonstrator} development
programme has followed two concurrent design directions.
The two designs, known as FE-A and FE-B, were aimed at the
DMILL and Honeywell radiation-hard process' respectively.
To date both have been realised in radiation-soft technologies,
(namely AMS and HP), and evaluated extensively in laboratory
and testbeam environments.
The FE-A design, (which was initially a BiCMOS design), was
also laid out and fabricated as a pure CMOS chip
which is referred to as the FE-C. 
The two design efforts have joined forces 
and are now on the threshold of the first radiation-hard submission
(to DMILL), of a common demonstrator design, (FE-D), which is largely based 
on combined features
of the FE-A and FE-B approaches. Later in 1999 it is planned 
that a common design submission also be made for the Honeywell
SOI process (FE-H).
All of the demonstrator chips were designed to be fully pin-compatible
enabling a common hardware test system to be used for all cases.
Only minor software and firmware options are required to switch between 
different chips.

There were several philosophical operational differences between 
FE-A and FE-B not least of which was the use of a dual-threshold discriminator 
for the FE-B front-ends as opposed to the faster single discriminator 
approach used in FE-A. 
Time information is derived from a lower
time-threshold whilst the upper threshold is resposible for hit-adjudication.
Both approaches incorporate negative polarity preamps for the 
$n^+$ in $n^-$ sensors, LVDS inputs and outputs for clock and control lines, 
a serial command protocol and serial readout 
scheme for line reduction and local data storage in the form of
end-of-column-pair buffer sets and a FIFO for hits which have matched
incoming triggers.
They also both incorporate eight 8-bit local DACs for front-end biases,
a local chopper circuit for calibration charge injection, 7-bits of digital
charge measurement based on time-over-threshold, channel-by-channel
masking capability for separate strobe and readout selection
and a 3-bit local threshold-tuning DAC, (TDAC), in every pixel cell 
for threshold-dispersion reduction.

\section{Results from demonstrator-assembly testing}
Throughout 1998 the collaboration followed a rich program
of testing of `single-chip assemblies' which comprised
the radiation-soft demonstrator electronics bump-bonded
to the first-generation prototype sensors, (both irradiated and
non-irradiated samples).
Testing has also progressed on a smaller number of full-scale modules 
with 16 FE readout-chips connected to a single tile.
\begin{figure}[tbp]	
\centerline{
\epsfxsize 8 cm 
\epsfbox{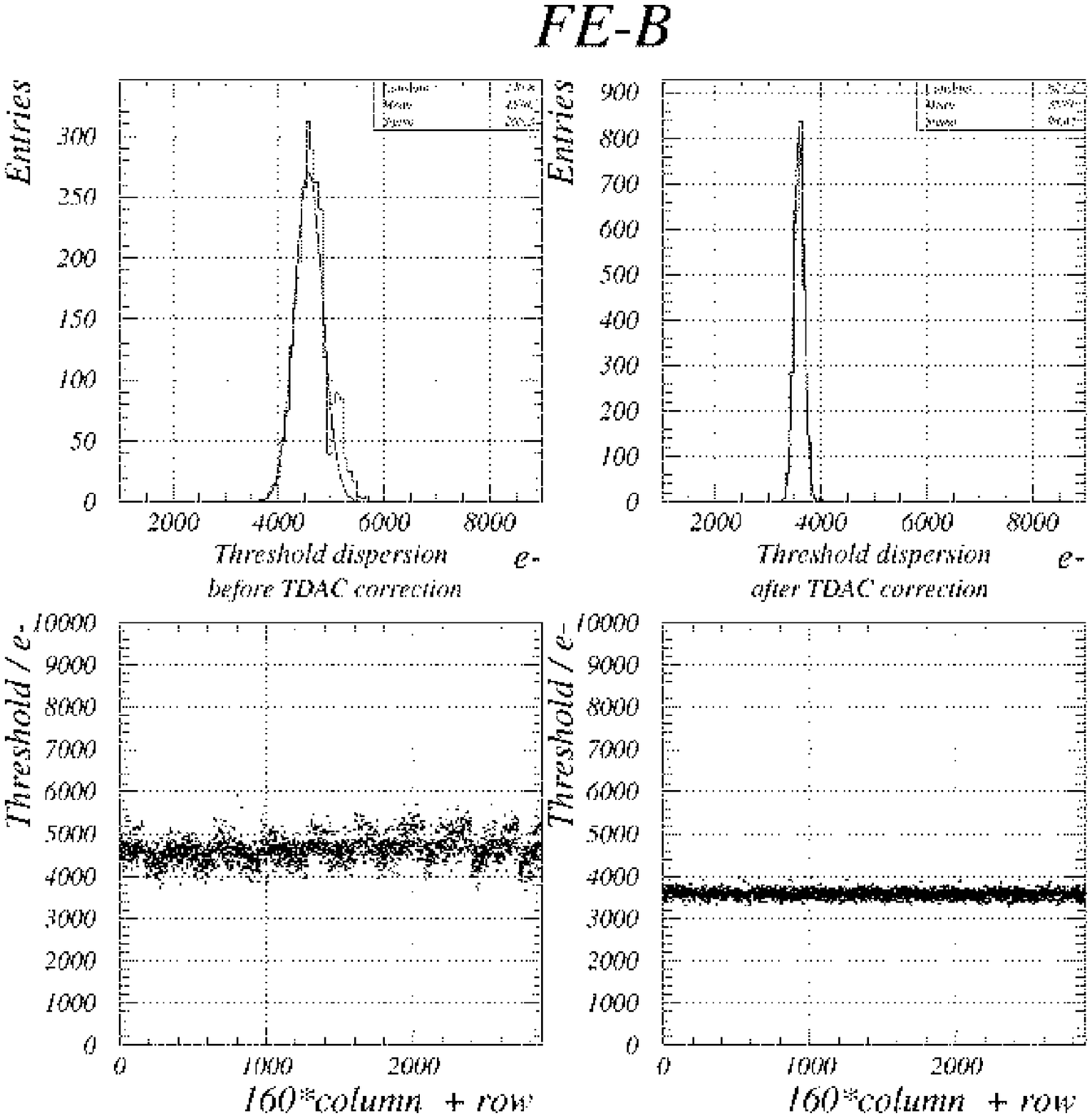}
\epsfxsize 8 cm 
\epsfbox{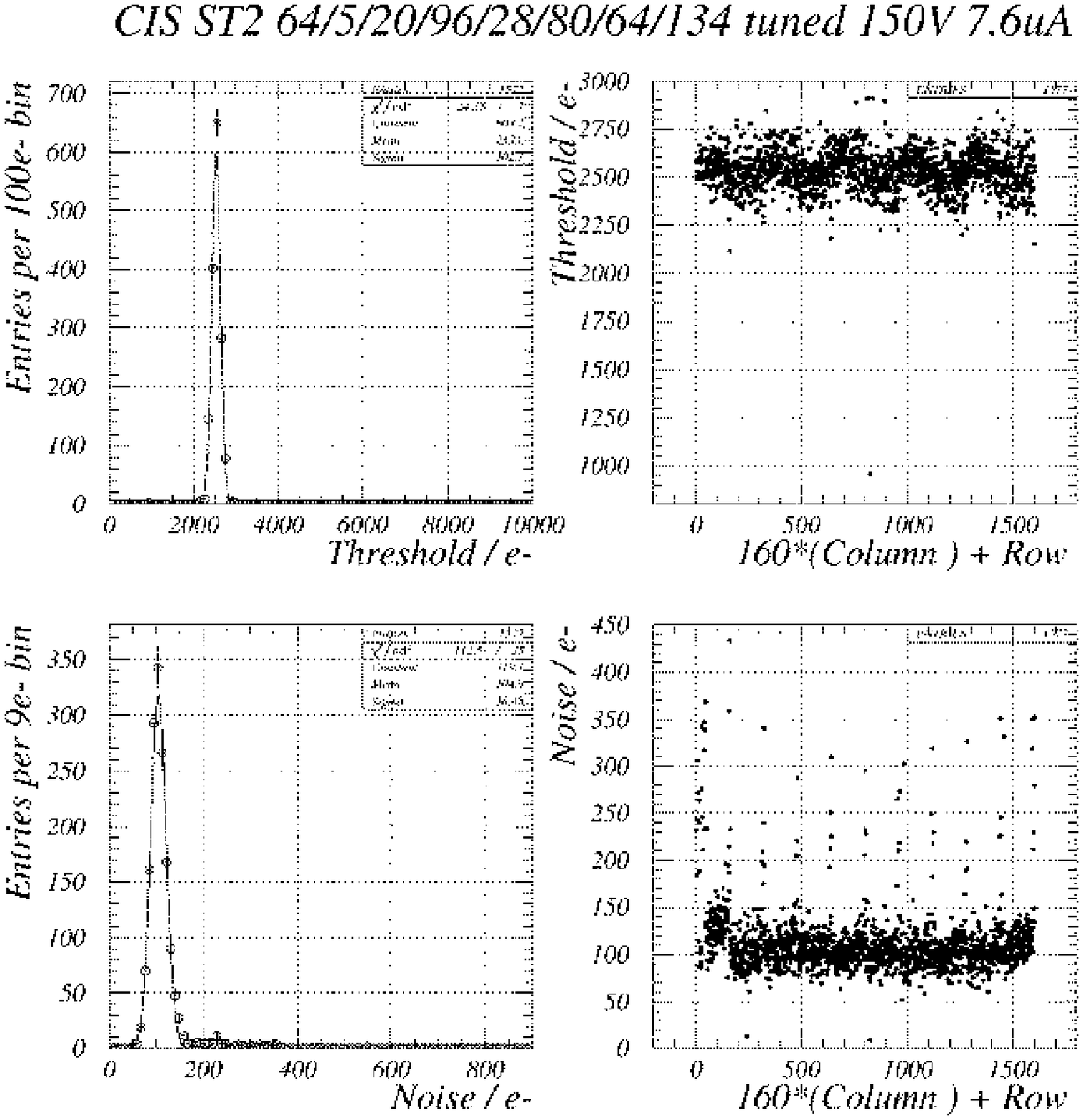}}
\vskip 0.3cm
\caption{\it Threshold dispersion before and after `TDAC'-tuning for a bare
FE chip, (left), and threshold and noise distributions after tuning for a single-chip assembly with an un-irradiated ST2 sensor. $\sigma_{Thresholds}$ = 103e-
and ENC = 105e-.}
\label{fi:tdactun}
\end{figure}
The top-left plot in figure \ref{fi:tdactun} shows the distribution
of thresholds for a bare chip. This is formed from scanning
the input calibration-pulse amplitude in fine steps and deriving
an `s-curve' of efficiency versus charge (knowing the
magnitude of the calibration capacitance) pixel-by-pixel.
The width of this distribution is 265e- with all of the 
3-bit TDACs set to the same value. On the right of this plot
the TDAC values for all channels have been assigned in order to 
minimise the dispersion. This process results in a post-tune distribution
width of 94e-.

The two lower plots on the left of figure \ref{fi:tdactun} show
the threshold values plotted according to `channel-number' which
is defined to be 160$\times$column number + row number. 
The four rightmost plots show the tuned-threshold distribution and
noise distribution for a typical FE-B single-chip assembly incorporating
an ST2 sensor. The equivalent noise charge (ENC) is measured to be
105e-. Similar results are obtained for ST1 assemblies whereas
the noise for an SSG device was measured to be 170e- due to the
higher load capacitance.

\begin{figure}[tbp]	
\centerline{
\epsfxsize 8 cm 
\epsfysize 8 cm
\epsfbox{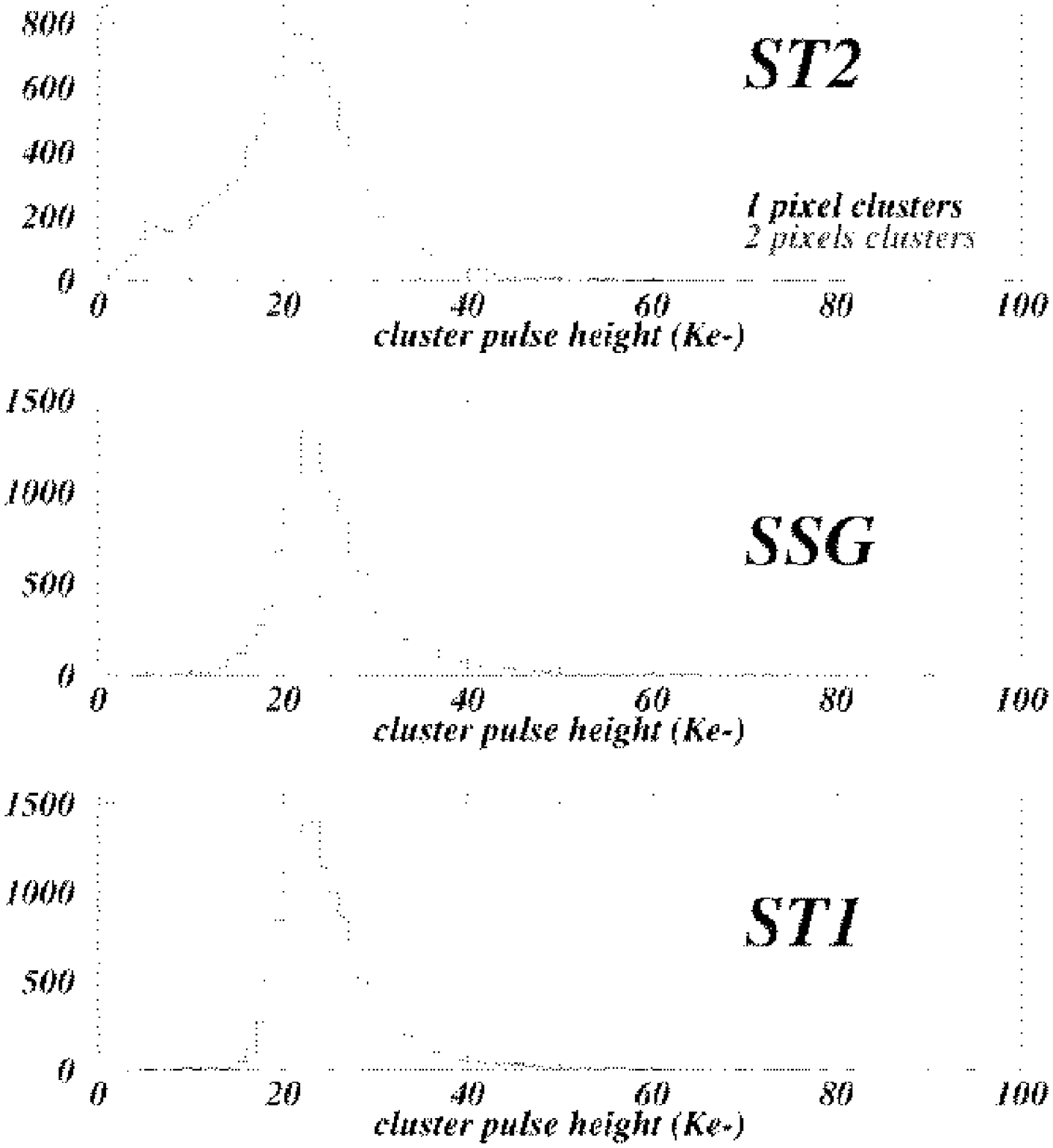}
\epsfxsize 8 cm
\epsfysize 8 cm 
\epsfbox{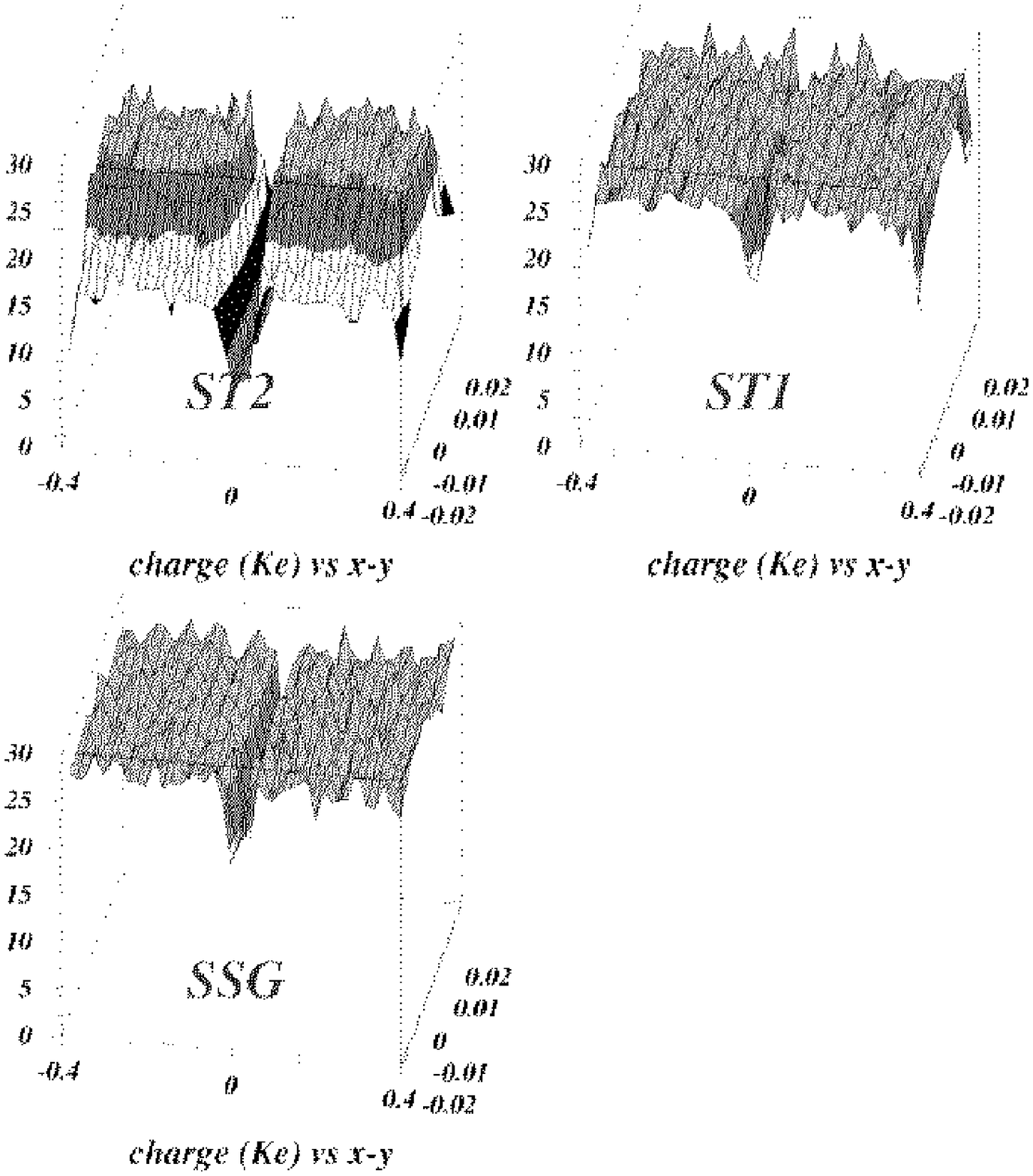}
}
\vskip 0.3cm
\caption{\it Testbeam results obtained for single-chip assemblies
incorporating the ST1, ST2 and SSG sensor design philosophies.}
\label{fi:h81}
\end{figure}

\begin{figure}[tbp]	
\centerline{
\epsfxsize 7 cm 
\epsfysize 6.2 cm
\epsfbox{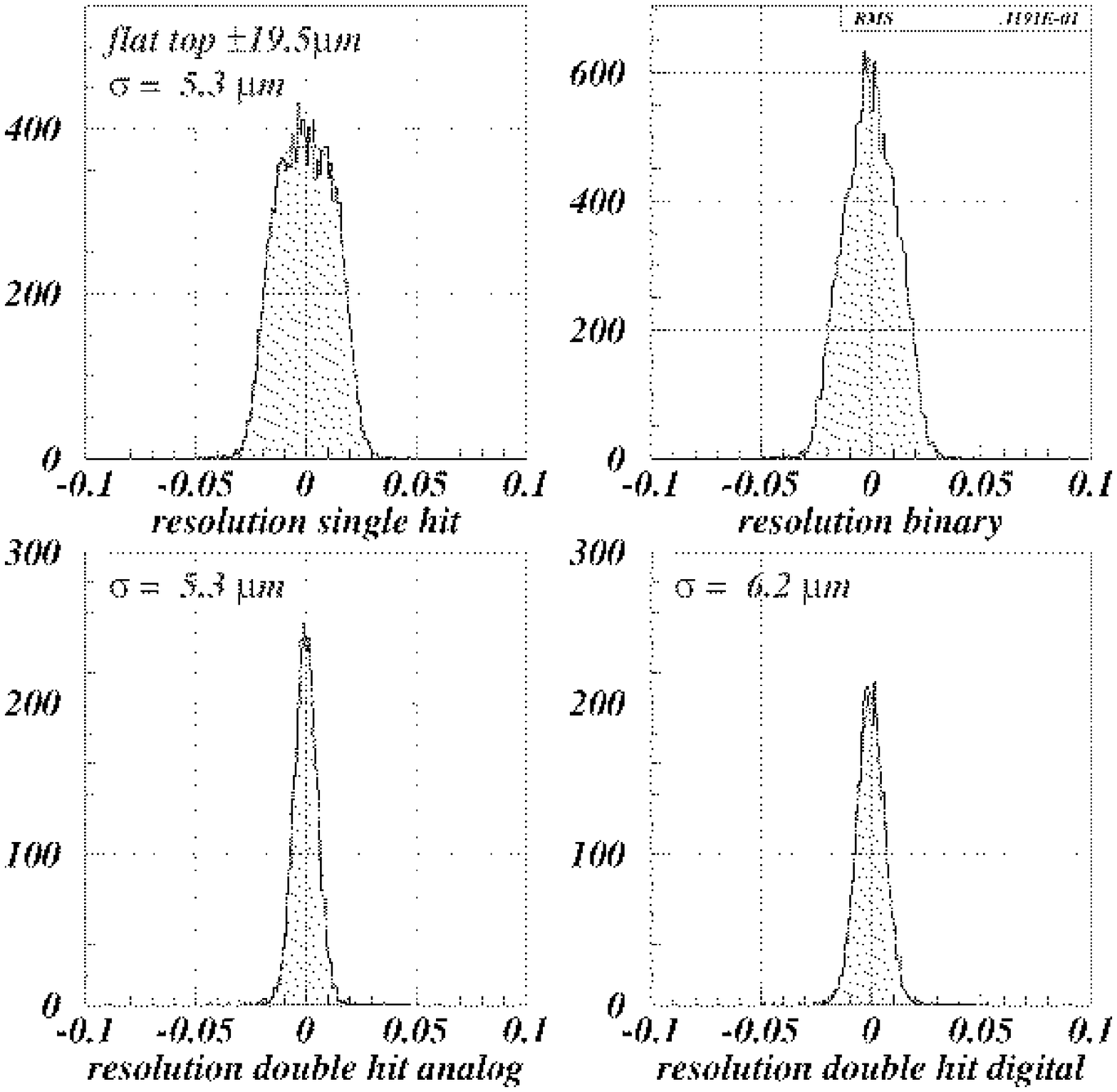}
\hspace{1cm}
\epsfxsize 7 cm 
\epsfysize 6.2 cm
\epsfbox{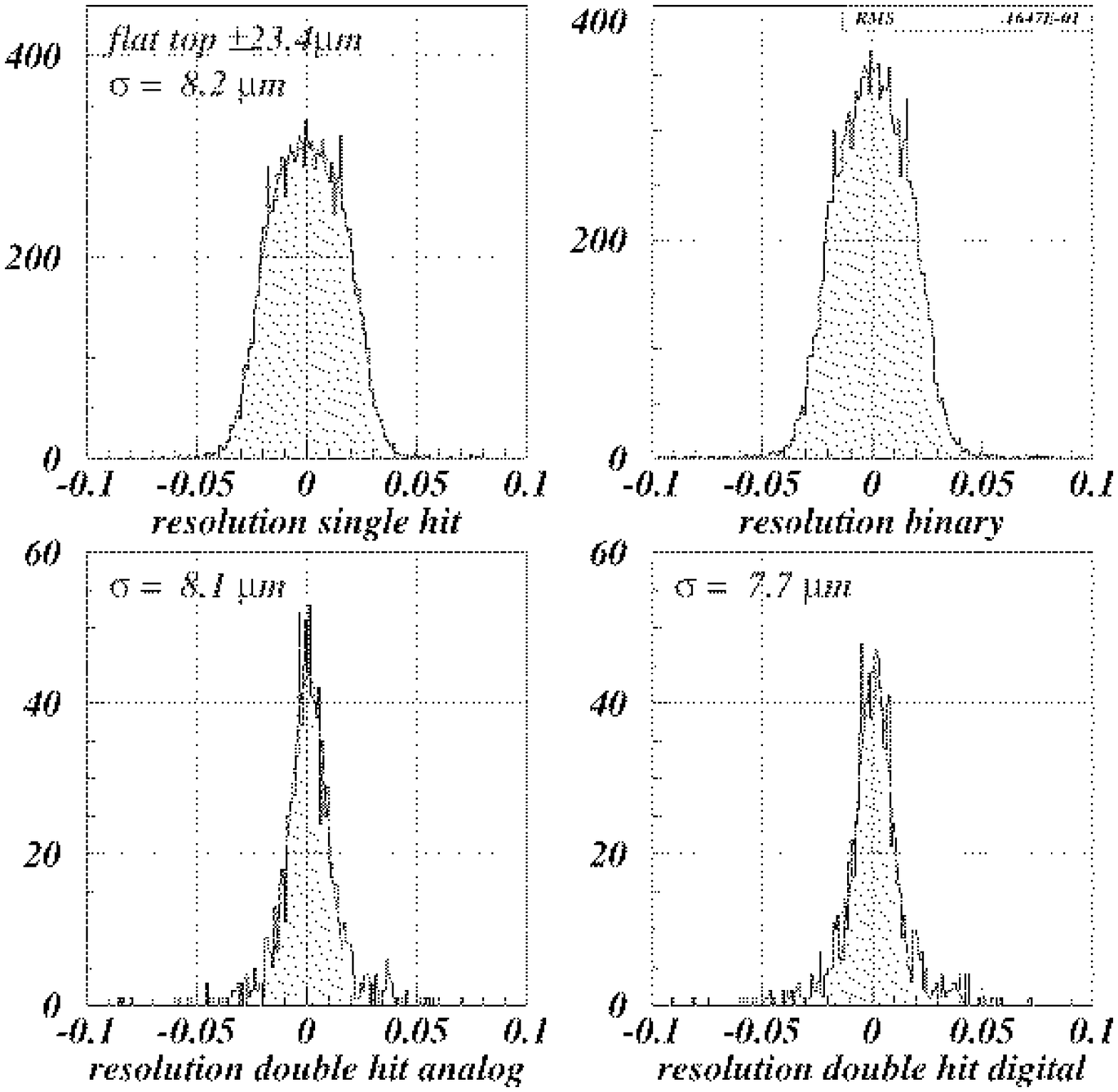}
}
\vskip 0.3cm
\caption{\it Residual distributions obtained for the SSG design (left) 
and for the ST2 sample irradiated to 
1.0$\times$10$^{15}n_{\mbox{eq}}$cm$^{-2}$
for tracks of normal incidence.}
\label{fi:h82}
\end{figure}

A range of assemblies were evaluated in the H8 testbeam of CERN's SPS
accelerator during four periods of 1998 running. 
The H8 facility offers a four cross-plane silicon microstrip
hodoscope, providing an extrapolation resolution of 3$\mu$m
for charged pions with momenta in excess of 100GeV/c.
This enabled detailed studies of hit-efficiency, charge-collection efficiency,
resolution and noise occupancy to form part of the comparative
process for the different prototype sensor designs both before
and after irradiation.
The left-most plots in figure \ref{fi:h81} show the charge spectra
obtained from the three principal design types, ST2, SSG and ST1 for
tracks of normal incidence.
Each plot has two histograms, one corresponding to the case where only a 
single channel was seen to fire and the other for double-channel clusters.
In all cases except for ST2, the spectra show the usual 
Landau/Gaussian-convolution shape, peaking at around the 21,000e- 
value expected for the 280$\mu$m thickness of Silicon.
For the ST2 case the double-channel distribution has a
distinctly lower peak value indicating some degree of charge-loss. 
In the central plots, a mean-charge surface is plotted
as a function of the local extrapolation point within a 2-pixel
cell. 
For the ST1 and SSG cases it is clear that there is a slight degree
of charge-collection inefficiency in the region between
neighbouring columns. For the ST2 design, the loss is extreme in this region
and further loss is apparent all around the pixel-cell outline.
This effect is directly attributable to the atoll $n^+$-rings which
surround each pixel in the ST2 design. The leftmost plots
in figure \ref{fi:h82} show residual distributions for an SSG assembly
for tracks of normal incidence. The upper right distribution
corresponds to the overall binary case in the $r\phi$ dimension which 
yields an r.m.s. of 12.9$\mu$m. This is slightly better than the expected 
50$\mu$m/$\sqrt{12}$ due to a small admixture of double-channel clusters.
Similar results were obtained for all of the designs pre-irradiation.

The table overleaf summarises the overall efficiency
figures obtained for the ST1, ST2 and SSG assemblies for 
thresholds of approximately 2ke- and 3ke-. 
The values are at least equal to 99\% in almost
all cases upon application of a range of quality cuts 
in time and space, the single exception being ST2 at 3ke- threshold
which registers 98.8\%. This is due to the worse charge-collection 
efficiency mentioned earlier.

Irradiated ST1 and ST2 samples were also tested in H8 at -10$^{\mbox{o}}$C.
ST2 samples which had received
5.0$\times$10$^{14}n_{\mbox{eq}}$cm$^{-2}$
and 
1.0$\times$10$^{15}n_{\mbox{eq}}$cm$^{-2}$ 
both performed extremely well for various reverse-bias voltages
up to 600V, as tabulated in the right-hand table below. 
An efficiency of 98.2\% was recorded for the maximum dose sample
when the analysis was restricted to a region within each pixel cell
which eliminates the charge-loss effects due to the atoll $n^+$ structures.
The fake-occupancy probability for this device 
was recorded to be 0.9$\times$10$^{-7}$ and half of those `hits' were
identified as originating from physics.
On the right of figure \ref{fi:h82} are the distributions of residuals
for the high-dose ST2 sample. The overall binary resolution is
measured to be 16.5$\mu$m.
The corresponding ST1 irradiated samples in contrast showed 
poorer performance. Extremely 
large numbers of noise hits were recorded for even the 
5$\times$10$^{14}n_{\mbox{eq}}$cm$^{-2}$ sample for 
applied detector biases in excess of 10V.

\begin{figure}[tbp]	
\centerline{
\epsfxsize 8.8 cm
\epsfysize 4 cm 
\epsfbox{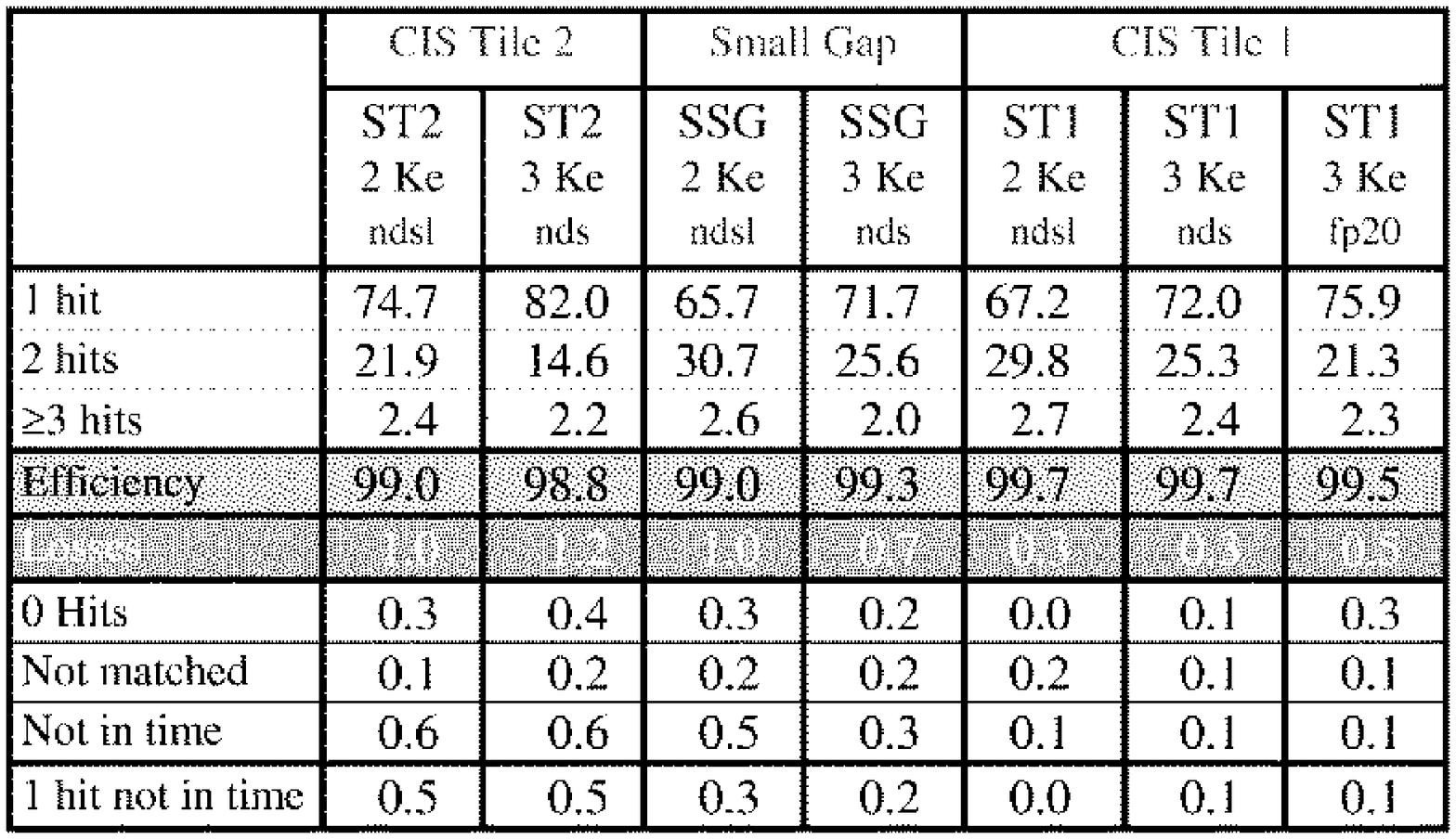}
\epsfxsize 8.8 cm 
\epsfysize 4 cm 
\epsfbox{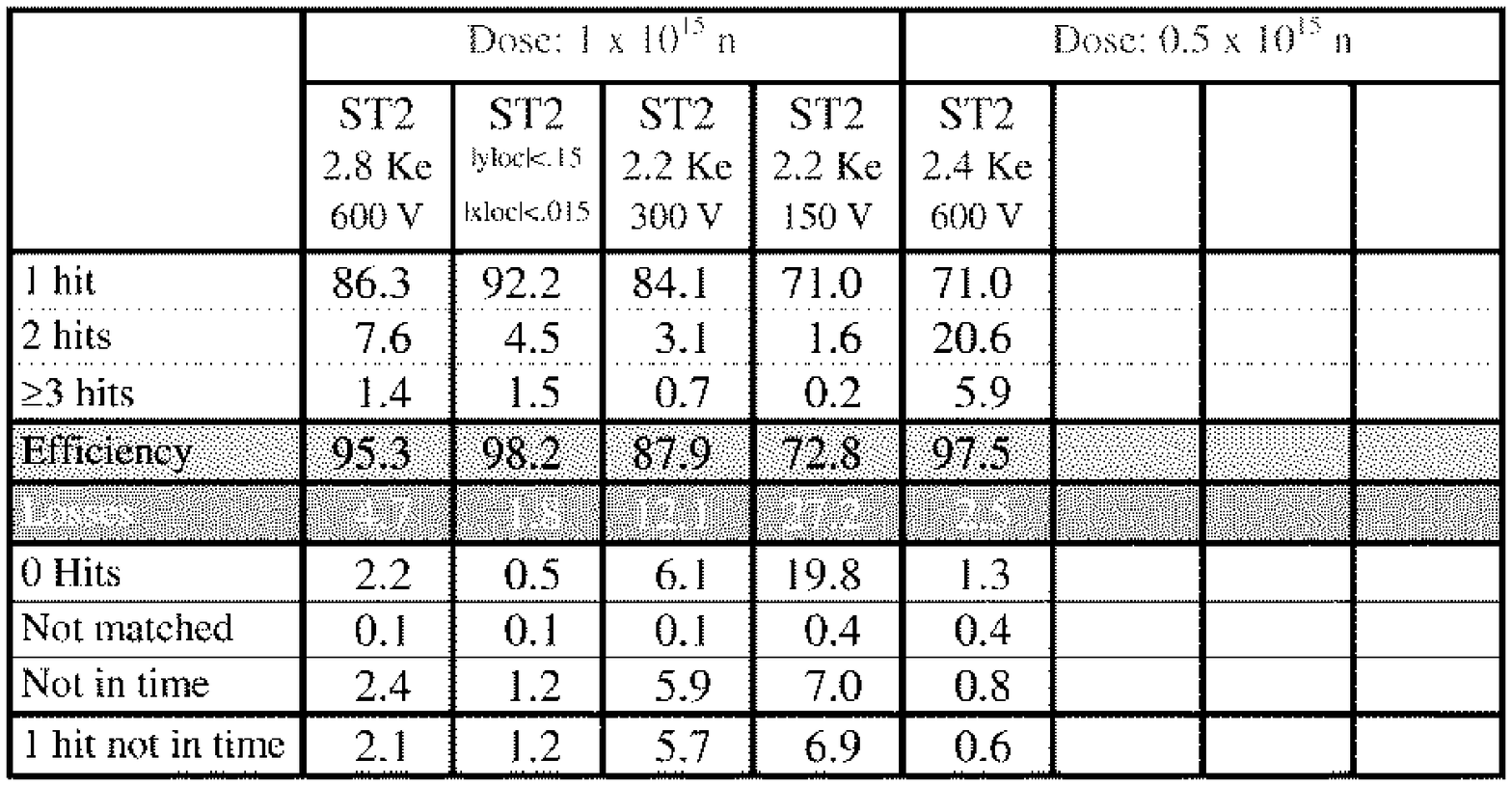}}
\vskip 0.3cm
\caption{\it Tables summarising the efficiencies of single-chip
assemblies pre-irradiation, (left), and after the half and full 
10-year hadronic
fluence in layer-1 for ST2, (right), under different operation conditions.
The sources of inefficiency are listed.}
\end{figure}
\begin{figure}[tbp]	
\centerline{
\epsfxsize 7 cm
\epsfysize 5.8 cm
\epsfbox{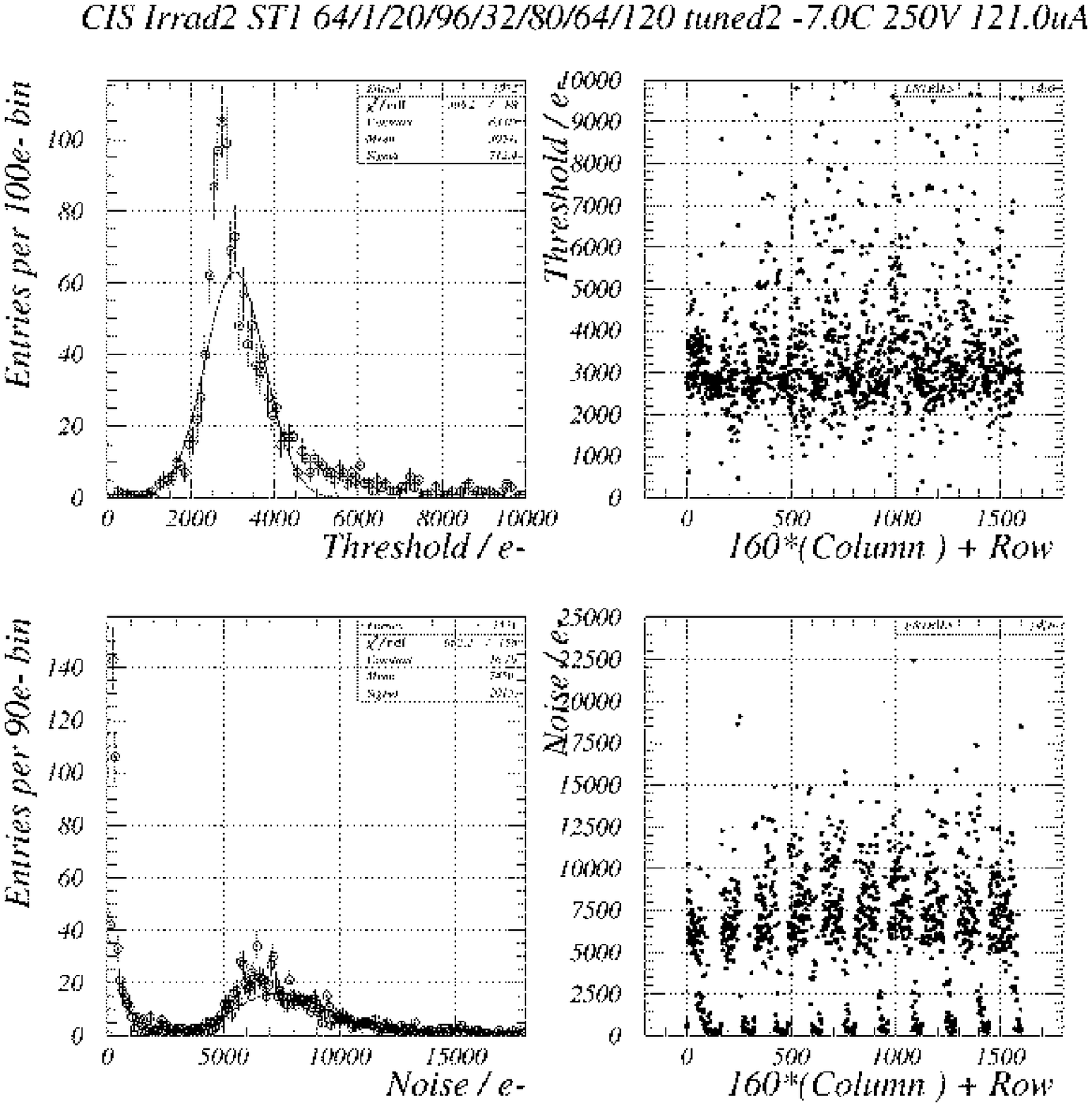}
\hspace{1cm}
\epsfxsize 7 cm 
\epsfysize 5.8 cm
\epsfbox{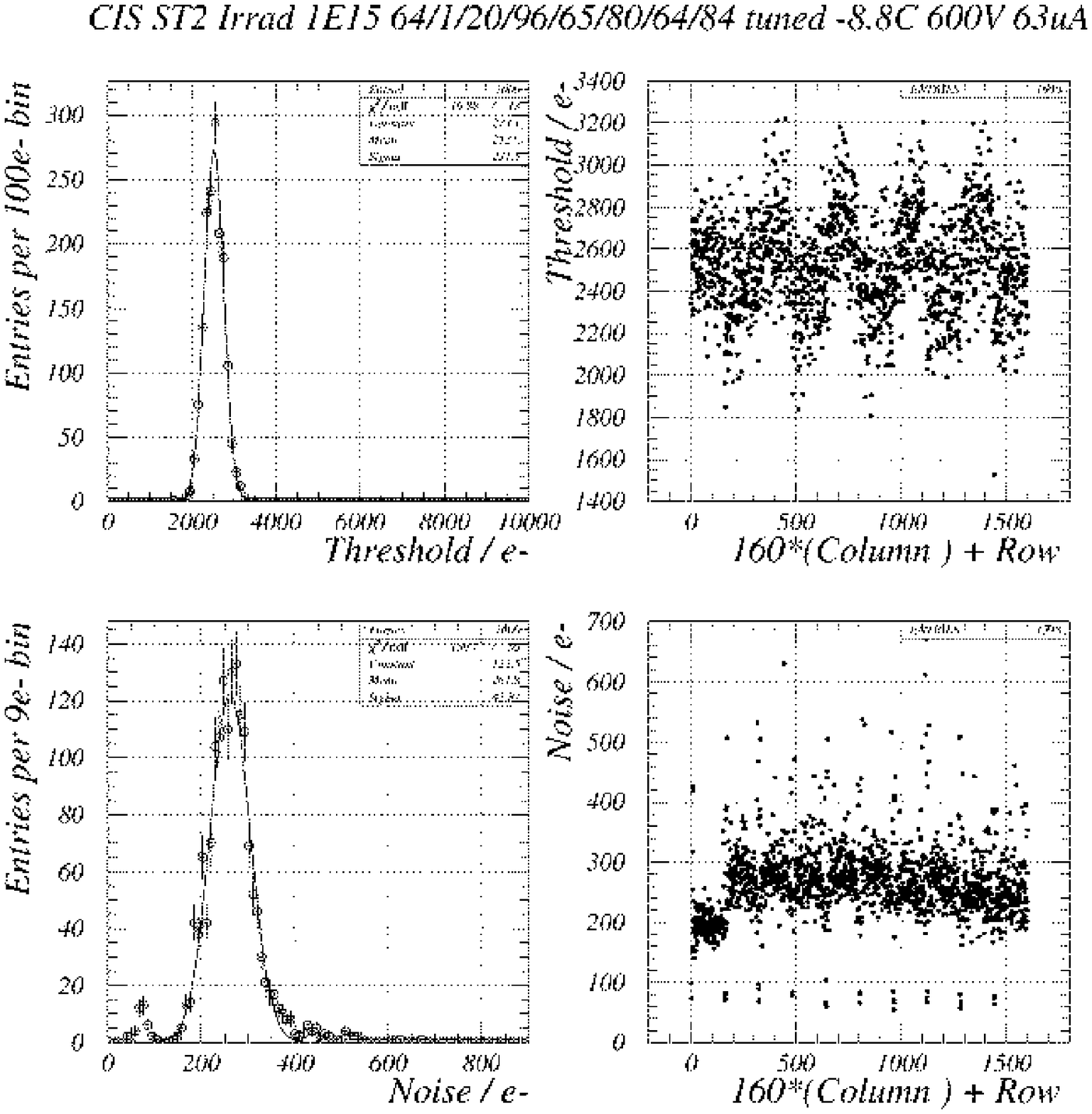}}
\vskip 0.3cm
\caption{\it Distribution of threshold and noise for the 
ST1 (left) and ST2 samples irradiated to 1.0$\times$10$^{15}n_{\mbox{eq}}$cm$^{-2}$ at 250V bias and 600V bias respectively.}
\label{fi:irrnoise}
\end{figure}

Noise analyses of the irradiated ST1 samples in the laboratory indicate
severe micro-discharge occurance once the backplane bias is
raised much above 100V. At 250V a broad spectrum of noise values
are recorded up to 15,000e-.
Even the highly dosed ST2 assembly on the other hand yields a very tight 
noise distribution at 600V in the laboratory with a mean of just 260e-,
(see figure \ref{fi:irrnoise}).

\begin{figure}[tbp]	
\centerline{
\epsfxsize 7 cm
\epsfysize 5.8 cm
\epsfbox{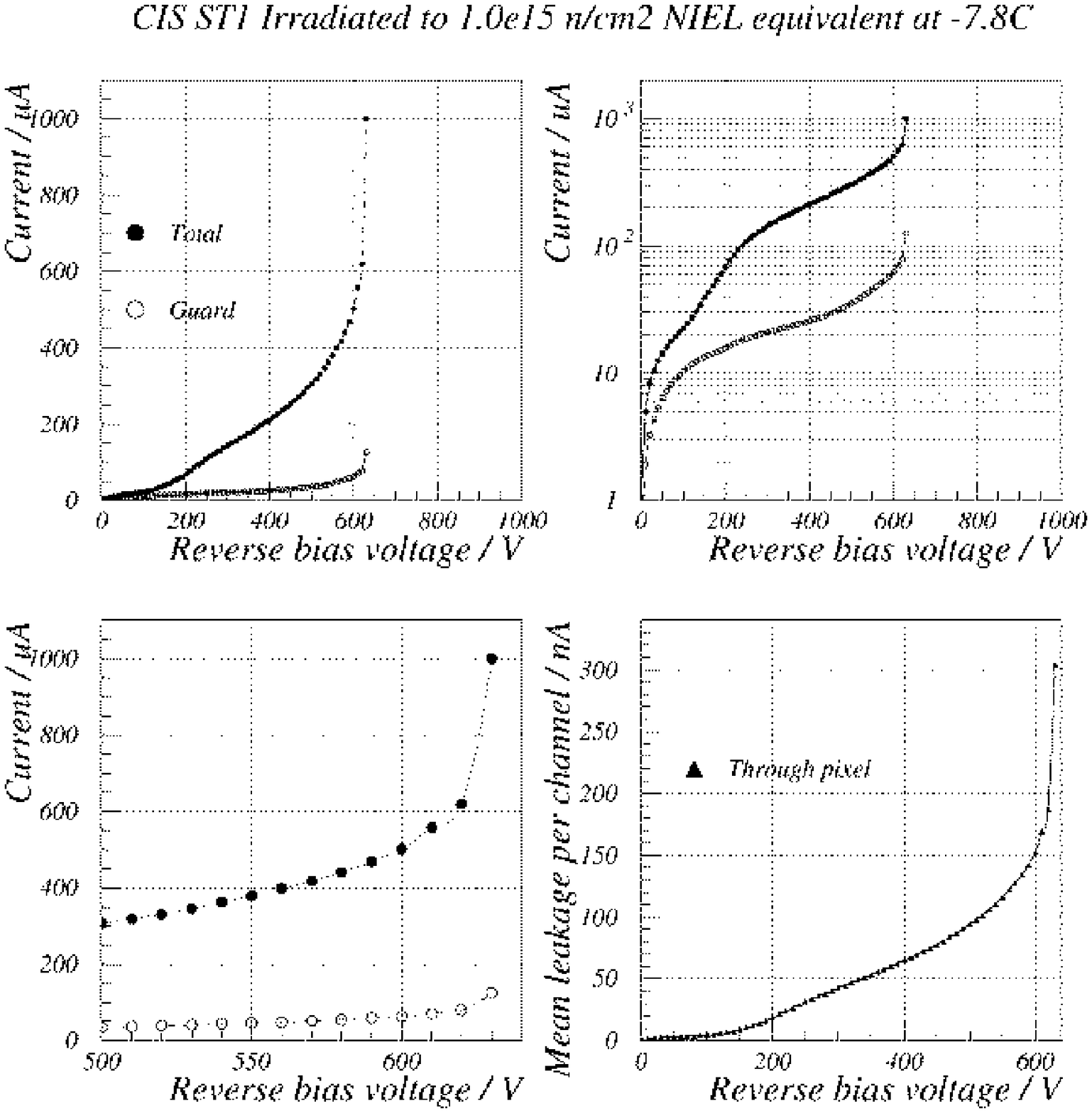}
\hspace{1cm}
\epsfxsize 7 cm 
\epsfysize 5.8 cm
\epsfbox{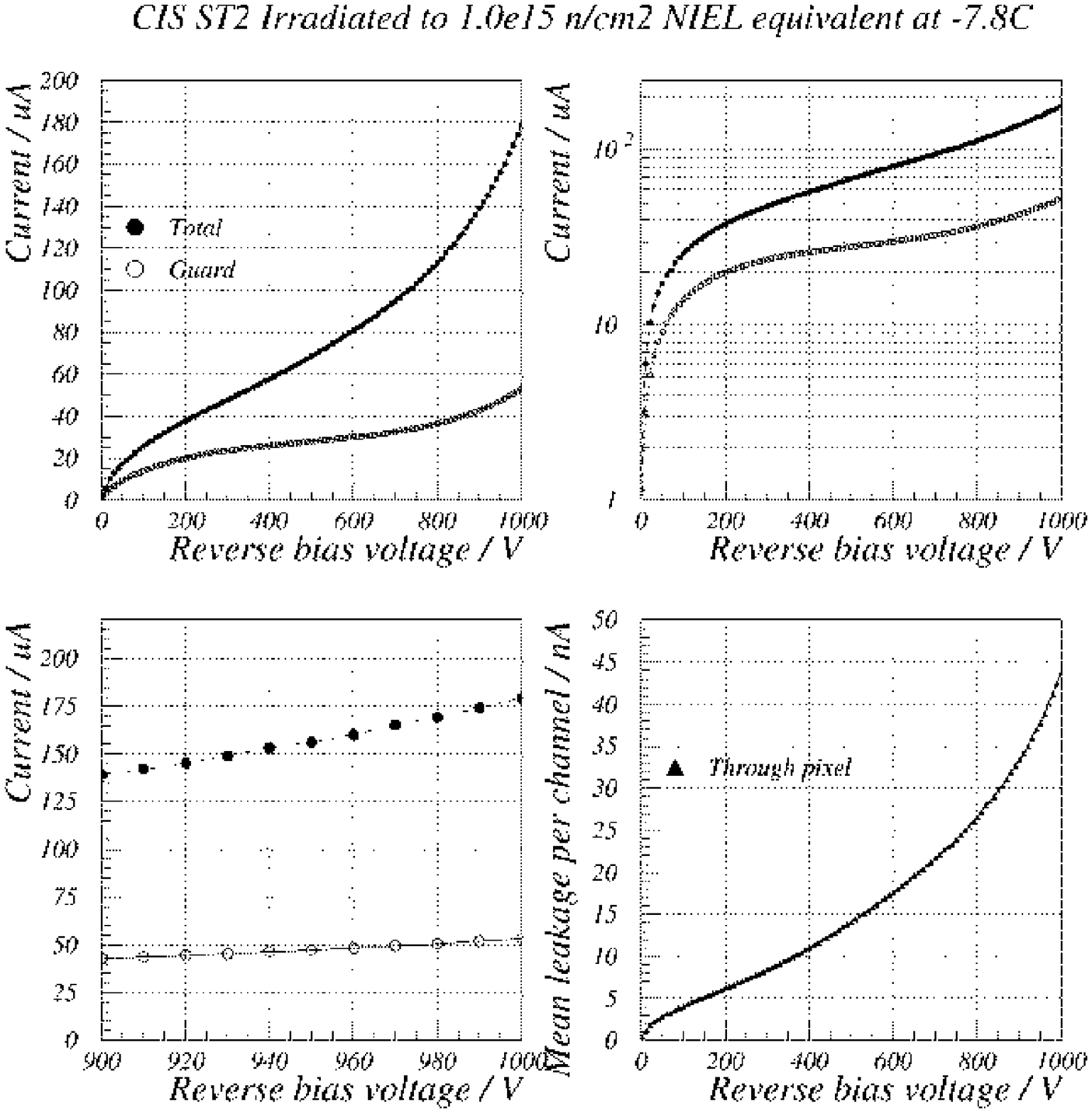}}
\vskip 0.3cm
\caption{\it I-V characterisations of the ST1 (left) and ST2 (right) sensors
irradiated to the pixel-layer-1 equivalent 10-year hadronic fluence
performed at $\approx$ -8$^{\mbox{o}}$C, up to 1000V reverse bias.
The ST1 device shows catastrophic breakdown at 630V whereas the ST2
device is electrostatically stable over the full range.}
\label{fi:bigiv}
\end{figure}

Figure \ref{fi:bigiv} shows the results of I-V characterisations
of the high-dose irradiated samples.
The ST1 assembly breaks down at an applied bias of around 630V whereas the ST2
device shows no signs of breakdown even up to 1kV.
All of these results strongly reinforce the view that p-spray isolation
leads to lower surface fields than the p-stop approach 
and thus greater micro-discharge and breakdown immunity 
when operated at high-voltage.
The next generation sensor prototypes reflect these observations
with the principal design based on p-spray isolation without 
any intermediate $n^+$ structures. This design looks like
the SSG approach except that the gaps are made slightly wider
in order to optimise the noise performance.
\section{Conclusions}
The ATLAS Pixel demonstrator program is progressing well 
with very encouraging results obtained from  
radiation-soft versions of the front-end readout electronics. 
The success of the electronics program has enabled   
evaluation of a series of prototype sensor designs
which have
provided the collaboration with a clear concept for a new
prototype layout.
This new design is optimised for high-voltage operation
post-irradiation
whilst maintaining good charge-collection efficiency.
The experience obtained with the two front-end design concepts
(FE-A and FE-B) has been of great benefit in converging on
combined designs for the upcoming radiation hard submissions.

\end{document}